\documentclass{icrc2009}

\usepackage{graphicx}   
\usepackage{caption}    
\usepackage[font=footnotesize]{subfig} 
\usepackage{fixltx2e}
\usepackage{url}

\newcommand{\shorttitle}[1]%
{\markboth{Proceedings of the 31\MakeLowercase{$^{st}$} ICRC, {\L}\'{o}d\'{z} 2009}{#1} }
\newcommand{\etal}{\MakeLowercase{\textit{et al. }}} 

\begin{document}
\title{Instrument simulation for the analysis of cosmic ray electron 
with the Fermi LAT}

\author{\IEEEauthorblockN{ C. Sgr\`o\IEEEauthorrefmark{1}, 
    J. Bregeon\IEEEauthorrefmark{1}
    and L. Baldini\IEEEauthorrefmark{1} \\ for the FERMI LAT collaboration} \\

\IEEEauthorblockA{\IEEEauthorrefmark{1}Istituto Nazionale di Fisica Nucleare, 
Sezione di Pisa, I-56127 Pisa, Italy}
}

\shorttitle{C. Sgr\`o \etal Simulation for Fermi LAT electrons}
\maketitle

\begin{abstract}
The Fermi LAT collaboration has built up a detailed Monte Carlo
simulation to characterize the instrument response and tune its
performance.
The simulation code is built around the widely used GEANT4 toolkit
and was carefully validated against beam test and flight data.
This poster shows how the full LAT simulation is used to
develop the event selection for the Cosmic-Ray Electron (\emph{CRE}\/)
analysis so as to optimize the instrument performance.
In particular, we will show in detail the determination of the geometry factor
and the residual hadron contamination.
The very accurate MC simulation proved to be fundamental
to control the systematic uncertainties
on the CRE spectrum measured by the Fermi LAT. 
  \end{abstract}

\begin{IEEEkeywords}
 Cosmic ray electrons; spectral analysis; Monte Carlo.
\end{IEEEkeywords}
 
\section{Introduction}
The Fermi Large Area Telescope (\emph{LAT}\/) is designed
to provide an accurate measurement of the incoming direction, 
energy and time of incident $\gamma$-rays that convert 
in $e^{+}+e^{-}$ pairs within the instrument. 
The LAT is composed of three detector subsystems. 
A tracker-converter made of silicon strip detector layers
interleaved with tungsten foils to enhance the photon conversion probability.
A calorimeter composed of CsI crystals arranged in a hodoscopic array 
in order to image the development of the electromagnetic shower 
and measure the energy of the incoming $\gamma$-ray.
A scintillator tiled anticoincidence detector 
that surrounds the tracker and provides the main identification of
charged vs. neutral incoming particles. 
More details on the instrument can be found in~\cite{LAT}.

Due to its capability to reconstruct electromagnetic showers 
and separate them from hadronic showers, the LAT is naturally a 
cosmic ray electron detector. 
No dedicated event reconstruction is required since the
topology of electron initiated showers is very similar to 
the one of photon initiated showers.

The main advantage of this instrument with respect to previous ones
is the large collection area and the long operation time 
that allows the accumulation of large
statistics during its life time. 
The main disadvantage is that electron-positron separation
is not possible\footnote{In this paper the word ``electron'' 
refers to the combination of positrons and electrons.}.

The event analysis to identify electron candidates 
as well as the evaluation of the instrument response functions 
is studied and optimized using a detailed Monte Carlo (\emph{MC}\/) simulation.
This paper describes the simulations required for the CRE study
and how they are used for the spectrum reconstruction~\cite{CRESpectrum}.

\section{Instrument Model}
The Fermi LAT colaboration uses a very detailed 
MC simulation to study the instrument
performance and the response to celestial sources. The software code
uses a precise description of the LAT geometry that includes the position
and the material of all the detector elements, both active and passive. 
The simulation is based on the GEANT4~\cite{GEANT4} toolkit, 
widely used in high-energy physics experiments, 
to model particle propagation and interaction through
the instrument (energy loss, multiple scattering, etc.) and therefore the energy
deposition in the active detector elements. A digitization algorithm
takes care of converting this information into raw detector quantities that
can be processed, by the reconstruction algorithms, in the same way of the
real data. The digitization algorithm 
has the capability of including detector ``defects'' like 
sensors misalignment, real gain and noise level, non linearities etc. 
In this way the reconstruction algorithm can use the actual calibration 
constants and the MC can provide a very accurate representation 
of the instrument response.

 \begin{figure}[!hbt]
  \centering
  \includegraphics[width=2.5in]{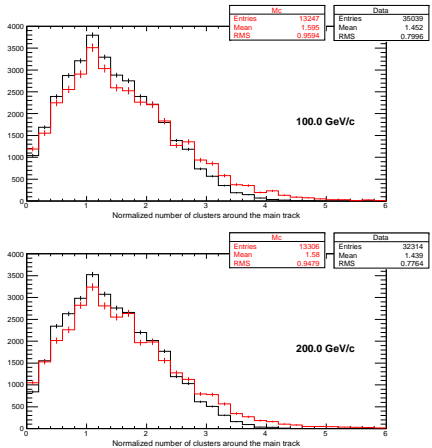}
  \caption{Number of hits around the best track
    (normalized to the track length) for a normal incidence electron beam 
    at $100$~GeV/c (top plot) and $200$~GeV/c (bottom plot).
    CU data in black and the corresponding MC simulation in red.}
  \label{BtVal}
 \end{figure}
The MC simulation has been extensively used for the design 
of the instrument and optimization of the event reconstruction 
and background rejection analysis.
The Instrument Response Functions that are used in the $\gamma$-ray analysis 
are evaluated using the MC. 
A similar scheme is employed in the CRE analysis and therefore the role of the 
simulation is crucial.

No Monte Carlo simulation is perfect, hence both
the physical processes and the geometry implementation
have to be validated by comparing the
simulation results with real data. The LAT MC has been verified on ground 
with cosmic rays and with a series of beam tests on a Calibration Unit 
(\emph{CU}\/) at CERN PS and SPS and at the GSI heavy ion 
accelerator laboratories~\cite{BT}. 
The Calibration Unit, composed of flight spare LAT modules 
(corresponding to about $1/8$ of the LAT), 
has been exposed to beams of electrons, hadrons and heavy ions 
in several configurations of beam energy and incoming angle. 
Figure~\ref{BtVal} shows an example of data-MC comparison 
for one of the tracker quantities that are relevant for CRE analysis.
The overall agreement between the MC simulations of the CU 
and the beam test data are good. 
A further verification of the agreement between LAT data and MC simulation
is performed using on-orbit simulation.

\section{On-orbit environment simulation}
The detection of the CRE with the LAT takes advantage of the on-orbit 
particle environment model built by the LAT collaboration.
This highly detailed model has been intensively used to develop all the
$\gamma$-ray background rejection algorithms, both on-orbit and off-line.
The model includes cosmic rays and earth albedo $\gamma$-rays 
starting from $10$ MeV and is valid outside
the South Atlantic Anomaly (since the LAT does not take data inside).
The particle fluxes are chosen to fit experimental data 
of several past experiments, more details can be found in~\cite{LAT}.
In addition to the primary component of the cosmic radiation,
the model includes the secondary particles produced by the interaction
with the atmosphere and the Earth magnetic field taking into account 
the geomagnetic cutoff and East-West effect.

\begin{figure}[!htb]
  \centering
  \includegraphics[width=2.5in]{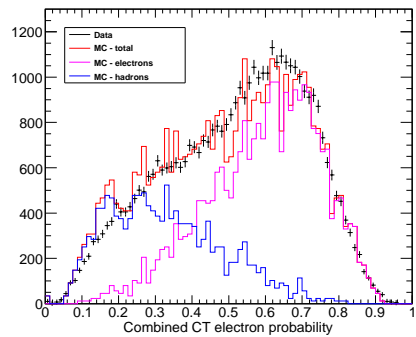}
  \caption{Electron probability from the Classification Tree output. 
    LAT data are compared with on orbit background model 
    for events above $100$~GeV. The selection on all the other quantities 
    is already applied.}
  \label{CT}
 \end{figure}

This model plays a fundamental role in the CRE analysis.
It provides a realistic simulation of the background particles, 
mainly protons, for the LAT on-orbit.
Together with the electron simulation described in section~\ref{allEle},
it has been used to identify the quantities,
in the event reconstruction, that are most sensitive to the difference 
between electromagnetic and hadronic shower topologies. 
The CRE event selection is, thus, based on the information provided by
the simulation of the LAT in the true orbital environment.
Notice that the CRE event selection 
(described in more details in~\cite{CREAnalysis}) 
uses a Classification Tree technique 
which was trained on MC simulation.
Figure~\ref{CT} shows the output of the Classification Trees on a sample of
electrons and hadrons simulated using this detailed cosmic ray model. 
The comparison with real LAT data shows a good agreement.

\begin{figure}[!hbt]
  \centering
  \includegraphics[width=2.5in]{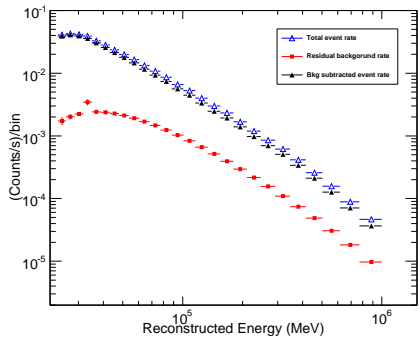}
  \caption{The residual background rate (in red) is evaluated, 
    for each energy bin, using the on-orbit environment simulation
    and is subtracted from the measured LAT event rate (in blue).
    The resulting rate (in black) is deconvolved with 
    instrument response functions to provide the electron spectrum.}
    \label{EvtRate}
 \end{figure}

The on-orbit background model was also used to evaluate and subtract 
the residual contamination in the electron event sample.
Because the model describes a realistic on-orbit environment,
it can be used to estimate the residual background rate 
as a function of the measured energy by dividing the
number of events that survive the event selection 
by the simulated collection time.
This estimate of the background rate is subtracted 
from the total event rate measured by the LAT as shown in figure~\ref{EvtRate}.
The residual contamination, evaluated as the ratio between 
residual background rate and the candidate CRE rate, 
provides a useful figure of merit for the quality 
of the analysis (fig.~\ref{GFCont}).
With this procedure the residual contamination depends 
only on the model of primary protons and its systematic uncertainty 
on the absolute scale is taken into account in the total systematic uncertainty.
The production of this MC sample was one of the most important 
and time consuming activities for the CRE spectrum measurement.
Because of the required high rejection power, 
about $10^3 - 10^4$ simulated events are needed to obtain one event 
that leaks through the selection and can be used for
residual background rate evaluation. 
The power law spectrum in $E^{-2.7}$ of primary protons (the main background source)
makes it even harder to populate the high energy region
of the spectrum.
To evaluate the residual background rate used in the CRE spectrum measurement,
about 400 CPUs for 80 days of computing time were used.

 
\section{Electron simulation}\label{allEle}
Together with the background model, a pure electron model
was used to simulate the LAT response to CRE.
This model does not include all the details of the LAT on-orbit,
but provides an isotropic flux in the LAT reference frame.
In this way the evaluation of the response functions does not
depend on orbital details while the isotropic distribution
of incoming electrons is a good assumption also for real operation.
The energy distribution in this model is a power law with index~$-1$,
so that we can accumulate high statistics in the whole energy range 
in a relatively short simulation time.\footnote{Notice 
that a $E^{-1}$ spectrum produces energy bins with equal number of events 
in case of logarithmic binning which is a common choice 
in high energy astrophysics}

\begin{figure}[!htb]
  \centering
  \includegraphics[width=2.5in]{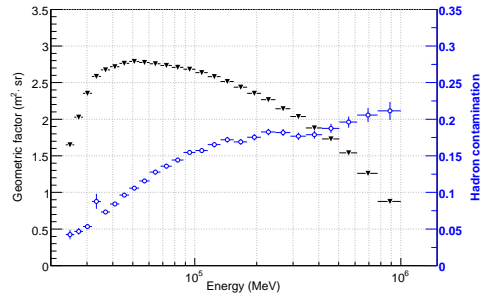}
  \caption{Geometry factor (in black) as a function of the true energy 
  and residual background contamination (in blue) as a function of reconstructed energy.}
  \label{GFCont}
 \end{figure}

The electron simulation was used to develop the event selection 
(together with the background model), but also to evaluate the LAT performance 
and the response functions for spectrum deconvolution.
These response functions parametrize the energy response of the LAT and its
geometrical acceptance as a function of true (MC) energy.
The acceptance (geometry factor) is calculated as proportional 
to the fraction of events surviving the selection in each energy bin.
The normalization constant depends on the area and the angle over which the
events have been generated.
The result of this calculation is shown in figure~\ref{GFCont}.
The sharp increase just after $30$~GeV is mainly due
to the on-board event selection designed to remove charged particles 
with deposited energy lower that $20$~GeV. 
The shape of the geometry factor is almost entirely due
to the event selection that must obtain a good background rejection power
in the whole energy range.

\begin{figure}[!htb]
  \centering
  \includegraphics[width=2.5in]{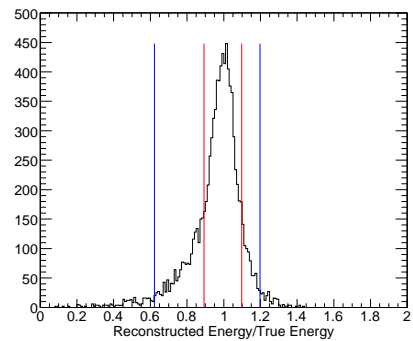}
  \caption{This figure shows how the energy resolution is evaluated 
    using the electron simulation. For each bin in true energy 
    (from $\sim 562$~GeV to $\sim 767$~GeV in this case) the distribution of 
    the reconstructed energy divided by the true energy is plotted.
    The vertical red lines show the boundaries of the $68\%$ containment window
    (for a resolution of $20\%$) and the blue lines are the boundaries of the 
    $95\%$ containment window.
  }
  \label{EnergyResCalc}
 \end{figure}

The energy dispersion is another important figure of merit
to quantify the LAT capability as an electron spectrometer.
It is evaluated dividing, event by event, the reconstructed
energy by the true energy. The resulting distribution peaks at $1$ 
and is asymmetric with a larger tail at low values 
(figure~\ref{EnergyResCalc}). 
The energy resolution is defined from this distribution 
as the full width of the smallest window that contains the $68\%$ of the events.
Usually also the $95\%$ containment is calculated in order to 
quantify the effect of the low energy tail.
The energy resolution, calculated in several bins of true energy,
is better than $10\%$ at $20$~GeV and get worse with the energy,
but without exceeding $30\%$ at $1$~TeV.
A further check on the energy reconstruction is done 
evaluating the most probable value by fitting the peak
with a LogNormal function and verifying that it does not 
differ significantly from $1$ in the whole energy range.
This ensures an unbiased response at least up $1$~TeV.

\section{Conclusions}
The Monte Carlo simulation has a central role 
in the measurement of the cosmic ray electron spectrum.
It was used to develop the electron selection algorithm
and to evaluate the instrument response functions for
spectrum reconstruction. 
The LAT collaboration made a large effort to build such a
detailed description of the instrument and its on-orbit behavior.
Also, the production of the large amount of simulated data required for the
electron analysis and in particular the evaluation of residual contamination
required an intense but successful work.
This extensive simulation, supported by beam test data 
and verified with flight data,
allowed the measurement of the most precise electron spectrum 
in the energy range $20$~GeV to $1$~TeV.

The $Fermi$ LAT Collaboration acknowledges support 
from a number of agencies and institutes for both development and the operation 
of the LAT as well as scientific data analysis. 
These include NASA and DOE in the United States, 
CEA/Irfu and IN2P3/CNRS in France, ASI and INFN in Italy, 
MEXT, KEK, and JAXA in Japan, and the K.~A.~Wallenberg Foundation, 
the Swedish Research Council and the National Space Board in Sweden. 
Additional support from INAF in Italy for science analysis 
during the operations phase is also gratefully acknowledged.


\newpage 

 \begin{thebibliography}{99}
 \bibitem{LAT}    W.~B.~Atwood et al., \emph{The Large Area Telescope on the Fermi Gamma-ray Space Telescope Mission},
   submitted to The Astrophysical Journal (arXiv:0902.1089v1)
 \bibitem{CRESpectrum}    A.~A.~Abdo et al., \emph{Measurement of the Cosmic Ray $e^{+}+e^{-}$ 
   spectrum from 20 GeV to 1 TeV with the Fermi Large Area Telescope},
   Phys.~Rev.~Lett., 102, (2009) 181101
 \bibitem{GEANT4} J.~Allison et al., \emph{Geant4 developments and applications},
   IEEE~Trans.~Nucl.~Sci., vol.  53 no. 1 (2006) 270-278
 \bibitem{BT}     L.~Baldini et al., \emph{Preliminary results of the LAT Calibration Unit beam tests}, 
   Proceedings of the first GLAST symposium, 
   AIP Conference Proceedings Vol. 921 (2007) 190-204              

 \bibitem{CREAnalysis} M.~N.~Mazziotta, \emph{Data analysis for the measurement of high energy cosmic ray electron/positron spectrum with Fermi-LAT}, 
   these proceedings 
      
 \end{thebibliography}
\end{document}